\documentclass[unsortedaddress]{revtex4-2}
\usepackage{paralist}
\usepackage{subfigure}
\usepackage{graphicx} 
\usepackage{xcolor}
\usepackage[shortcuts]{extdash}
\PassOptionsToPackage{hyphens}{url}\usepackage{hyperref}
\hypersetup{
    colorlinks=true,
    linkcolor=blue,
    citecolor=blue,
    filecolor=magenta,
    urlcolor=blue,
}

\usepackage{siunitx}

\begin{document}
\title{Materials Design for the Synthesis of High Strength Radiopure Copper Alloys for Rare Event Detection}

\author{Dimitra~Spathara}
\email[e-mail:]{d.spathara@bham.ac.uk}
\affiliation{School of Physics and Astronomy, University of Birmingham, Birmingham, B15 2TT, UK}
\author{Patrick~Knights}
\affiliation{School of Physics and Astronomy, University of Birmingham, Birmingham, B15 2TT, UK}
\author{Konstantinos~Nikolopoulos}
\affiliation{School of Physics and Astronomy, University of Birmingham, Birmingham, B15 2TT, UK}
\affiliation{Institute for Experimental Physics, University of Hamburg, Hamburg, 22761, Germany}

\date{\today}

\begin{abstract}
Additive-free electroformed copper has emerged as the material of choice in exceptionally radiopure detectors for rare-event searches, based on its radiopurity, physical properties, and affordability. However, copper is ductile and of limited mechanical strength posing challenges for its use in future experiments. Electroformed copper-based alloys have been identified as a promising solution. However, their synthesis needs refining by exploring a complex parameter space of compositions and strengthening mechanisms. Here we show how a materials design approach may address current challenges and optimize alloy synthesis and processing. Alloy properties are predicted following thermal processing, using computational thermodynamics. The findings suggest a methodology to design high-performance, radiopure  copper-based alloys suitable for next-generation rare-event experiments, while minimizing lengthy and expensive trial-and-error approaches. The impact on future experiments is exemplified through case-studies of the DarkSPHERE and XLZD experiments.
\end{abstract}


\maketitle

\section{Introduction}
Copper is widely used in detector systems for rare-event search experiments~\cite{Armengaud:2009hc,Auger:2012gs,CUORE:2017tlq,NEWS-G:2024jms} owing to its exceptional radiopurity, the absence of long-lived isotopes, excellent electrical and thermal properties, and commercial availability with purities up to 99.9999\%. To further minimize radioactive backgrounds, additive-free electroformed copper (EFCu)~\cite{hoppe2008use, Hoppe:2014nva} has been developed and optimized, using a variety of techniques~\cite{CHANDRASEKAR20083313}. As a result, it prominently features in several state-of-the-art detector systems~\cite{MAJORANA:2016lsk,NEWS-G:2020fhm} and its use is explored by many rare-event search experiments~\cite{Hoppe2022}. Electroforming underground further suppresses material activation from cosmic-ray-induced neutrons.

Despite these advantages, the high ductility and relatively low strength of Cu limit its application, particularly in moving, high-pressure, and load-bearing components. This led experiments, such as LZ~\cite{LZ:2019sgr} and DarkSide-20k~\cite{DarkSide-20k:2017zyg}, to use alternative structural materials, such as titanium or steel, to meet the requirements. Although these materials can be procured at high purity, the increasing sensitivity of rare-event search experiments means that these alternative materials have become a limiting factor in terms of background contributions. For example, it is estimated that the dominant background in the LZ experiment comes from the emanation of $^{222}$Rn from the Ti vessel and the steel piping~\cite{LZ:2022ysc}. Although this does not impact the scientific goals of LZ, it could be detrimental to a next-generation experiment, such as XLZD~\cite{XLZD:2024pdv}. 
Thus, in view of future rare-event search experiments, the development of materials with enhanced properties and radiopurity is extremely timely and important.

To address the mechanical limitations of Cu while benefiting from the extreme radiopurity of EFCu, early investigations have been carried out on the synthesis of Cu-based alloys that offer improved properties without compromising radiopurity~\cite{Suriano:2018nrb,Vitale:2021xrm}. 
Cu-based alloys are important for a wide range of applications such as transportation, electronics, and electric circuits~\cite{yang2023recent, peng2005property, yuan2017microstructure, xu2018effect}, for which the electrical conductivity of Cu is crucial and superior mechanical performance is required. The most promising alloying element for the improved strength of EFCu is chromium, where even small additions 
can substantially enhance the strength of the alloy~\cite{HUANG2021102378}. 
Initial observations indicate that Cr concentrations in the range of 0.3-0.58~wt\%, combined with heat treatment and aging, can significantly increase the hardness of EFCu by 70-100\%~\cite{Suriano:2018nrb,Vitale:2021xrm,osti_1039850}. 
Moreover, in Ref.~\cite{Suriano:2018nrb} the projected radiopurity of the CuCr alloy with 0.585~wt\% is at levels comparable to those of EFCu. Nevertheless, introducing Cr without compromising radiopurity is non-trivial, and the technique used to synthesize these CuCr alloys needs to be refined, by exploring a complex parameter space of compositions and strengthening mechanisms, to optimize the balance between strength and radiopurity.

Copper can be electrodeposited at relatively low potentials, around 0.3~V, which is advantageous because it effectively minimizes the co-deposition of thorium and uranium ions from the electrolyte and helps maintain high radiopurity. In contrast, Cr deposition requires higher voltages, typically $0.5\;\si{\volt}$.
Moreover, co-electrodeposition of Cu and Cr is challenging due to their different electrochemical behaviors~\cite{Vitale:2021xrm}. To minimize radioactive contamination, both Cu and Cr must be electroplated from additive-free, water-based solutions. Additives would favorably improve plating parameters, however, they introduce unwanted radioactive contamination. 
Therefore, initial studies have used a sequential deposition approach, in which thin layers of Cr and Cu are electrodeposited from separate baths, followed by post-deposition heat treatments to promote alloying and further strengthening~\cite{Suriano:2018nrb, Vitale:2021xrm}. 
In Ref.~\cite{Vitale:2021xrm}, specifically, thin Cr layers were electrodeposited on each side of EFCu, followed by a solution heat treatment and aging. 

In recent work, a computational thermodynamics-based methodology was developed and validated against the available measurements~\cite{Spathara:2025bfw}. Subsequently, it was applied to
optimize the layer configuration indicating 
that when $1\;\si{\micro\meter}$~Cr is electroplated on $400\;\si{\micro\meter}$ thick EFCu, a homogenized alloy composition of 0.2~wt\% Cr content is obtained after 24 h heat treatment at $1050\,^{\circ}$C. It is also understood that a Cr thickness of $10\;\si{\micro\meter}$ cannot be incorporated into a $400\;\si{\micro\meter}$ thick EFCu layer, due to the limited solubility of Cr in Cu~\cite{Vitale:2021xrm,Spathara:2025bfw}.

Furthermore, small additions of Ti have been found to improve the mechanical properties of the Cu-0.5Cr (in wt\%) alloy as-cast~\cite{HUANG2021102378}. Ti atoms have been reported to be incorporated into the faced centered cubic (fcc) matrix and inhibit further increase in Cr-rich spherical precipitates, which grow during the aging stage that follows the solution heat treatment stage. For CuCrTi alloys in the as-cast state, additions of 0.2 to 0.4 wt\% lead to a hardness increase at levels comparable to those of a mild steel, i.e. 160 to 220~HV (Vickers Hardness). Because alloy fabrication from the melt poses purity contamination risks, we explore the feasibility of incorporating trace amounts of Ti towards an electroformed CuCrTi alloy. An additional benefit of trace Ti additions, in contrast to those explored in Ref.~\cite{HUANG2021102378}, is that a high electrical conductivity is maintained~\cite{EZE2018163,yang2023recent}. In Ref.~\cite{EZE2018163}, trace Ti additions to Cu are explored and improved mechanical properties are observed for alloy compositions of Cu-0.014Ti and Cu-0.035Ti in wt\%. Finally, electroplating Ti requires non-aqueous solution and furthermore it is challenging to achieve without the use of additives. For this reason, in the following the CuCrTi alloy is designed around a pure Ti substrate.

In this article, we investigate materials for rare-event search experiments from a materials design perspective and explore their impact on the physics sensitivity of the experiments. We introduce a methodology based on computational thermodynamics~\cite{Spathara:2025bfw} to model and optimize the composition and processing parameters of high-strength, radiopure CuCr alloys. Building on CALPHAD-type~\cite{lukas2007computational, SPENCER20081, kattner2016calphad} thermodynamic and kinetic databases, we simulate the behavior of electrodeposited Cu and Cr layers during thermal processing, enabling the prediction of phase stability, transformations, and microstructural evolution. In order to achieve a homogeneous alloy composition of Cu-0.5Cr (in wt\%), we propose appropriate Cr/Cu layer configurations in the electroplating stage. This is a systematic application of computational thermodynamics to materials design for rare-event searches, which has not been previously reported in the literature. 
Moreover, we explore the feasibility and impact of the addition of trace Ti, which has been found to improve the mechanical properties of the Cu-0.5Cr (in wt\%) alloy~\cite{HUANG2021102378}, and appropriate configurations and heat treatments are proposed. The utilization of CuCrTi alloys has not been considered in the experimental particle and astro-particle physics literature.
The deployment of these cutting-edge tools to the development of high-radiopurity, high-strength alloys for rare-event search experiments has the potential to substantially reduce the development time required for these application-specific alloys. Their availability could enable experiments to achieve their targeted sensitivity and facilitate construction. The impact to future experimental sensitivity is demonstrated through two case studies of next-generation dark matter direct detection experiments. We opt for the DarkSPHERE~\cite{NEWS-G:2023qwh} and XLZD~\cite{XLZD:2024nsu} detectors, as they have complementary physics objectives and employ vastly different technologies. It is expected that the majority of future rare-event search experiments would benefit from the proposed developments.

\section{Results and Discussion}
\subsection{Manufacturing electroformed CuCr and CuCrTi alloys}

For the manufacturing of electroformed alloys, thermal processing, consisting of solution heat treatment and aging, is key. These two stages of thermal processing aim to achieve alloy homogenization and precipitation strengthening to achieve maximum mechanical enhancement, respectively, and need to be achieved without compromising radiopurity. Thus, optimization of the respective process parameters is required. Recently, these processing stages were successfully modeled, for example the nucleation and growth of precipitates in CuCrTi alloys as-cast were simulated~\cite{HUANG2021102378}, and were implemented in a methodology to achieve homogenized CuCr alloy compositions with single-phase microstructures~\cite{Spathara:2025bfw}.

In the following, the focus is on understanding the solution heat treatment stage. This is crucial to achieve a homogenized alloy composition, which is a prerequisite for any subsequent aging stage. Possible residual inhomogeneities of the alloy composition, could result in different sizes of precipitates following aging, and, thus, to local variations of the materials mechanical properties. The sensitivity to such fluctuations needs to be judged on an application-to-application basis. In the following, the aim is to achieve uniformity at the level of a few parts per thousand.

The Cr/Cu configuration initially consists of two regions of the corresponding pure elements, separated by a planar boundary. 
From simulations in Ref.~\cite{Spathara:2025bfw}, it is understood that homogenized alloy composition for the Cr/Cu layer configuration examined in Ref.~\cite{Vitale:2021xrm} could be achieved -- within practical time scales -- at $1050\,^{\circ}$C, in contrast to the  $1025\,^{\circ}$C used experimentally in Ref.~\cite{Vitale:2021xrm}. The pure Cr and Cu layers are found in body centered cubic (bcc) and fcc phases, respectively, at $1050\,^{\circ}$C. During solution heat treatment, the phase transformation from bcc to fcc occurs to a single-phase microstructure.

\subsubsection{Single and double-sided layer configuration for Cu-0.5Cr alloy}
\label{sec:alloy1}
Solution heat treatments at $1050\,^{\circ}$C have been simulated to determine the Cr/Cu layer configuration that would lead to a homogeneous alloy composition of Cu-0.5Cr (in wt\%). Examples of Cu and Cr layer thickness configurations that yield the desired properties are shown in Table~\ref{tab:config}.

\begin{table}[h!]
    \centering
    \caption{Layer configurations. Cr and Cu layer configurations for homogenized Cu-0.5Cr (in wt\%) alloy composition after solution heat treatment at $1050\,^{\circ}$C.\label{tab:config}}
    \begin{tabular}{c|ccc|cc} \hline
  Layer &       \multicolumn{3}{c|}{Thickness} & & \\
        Configuration & Cr layer &  Cu layer & Alloy  & Time & Figure\\
        & ($\si{\micro\meter}$) & ($\si{\micro\meter}$) & ($\si{\micro\meter}$) & &\\\hline
       Single-sided & 1.5 & 245 & 246.5 & 32~h & \ref{fig:1a}\\ 
       Double-sided & 3 & 245 & 493 & 32~h & \ref{fig:1b}\\
       Double-sided & 6 & 490 & 986 & 5~d & \ref{fig:1c} \\ \hline
    \end{tabular}
\end{table}

\begin{figure}
    \centering
    \subfigure[\label{fig:1a}]{\includegraphics[width=0.35\linewidth]{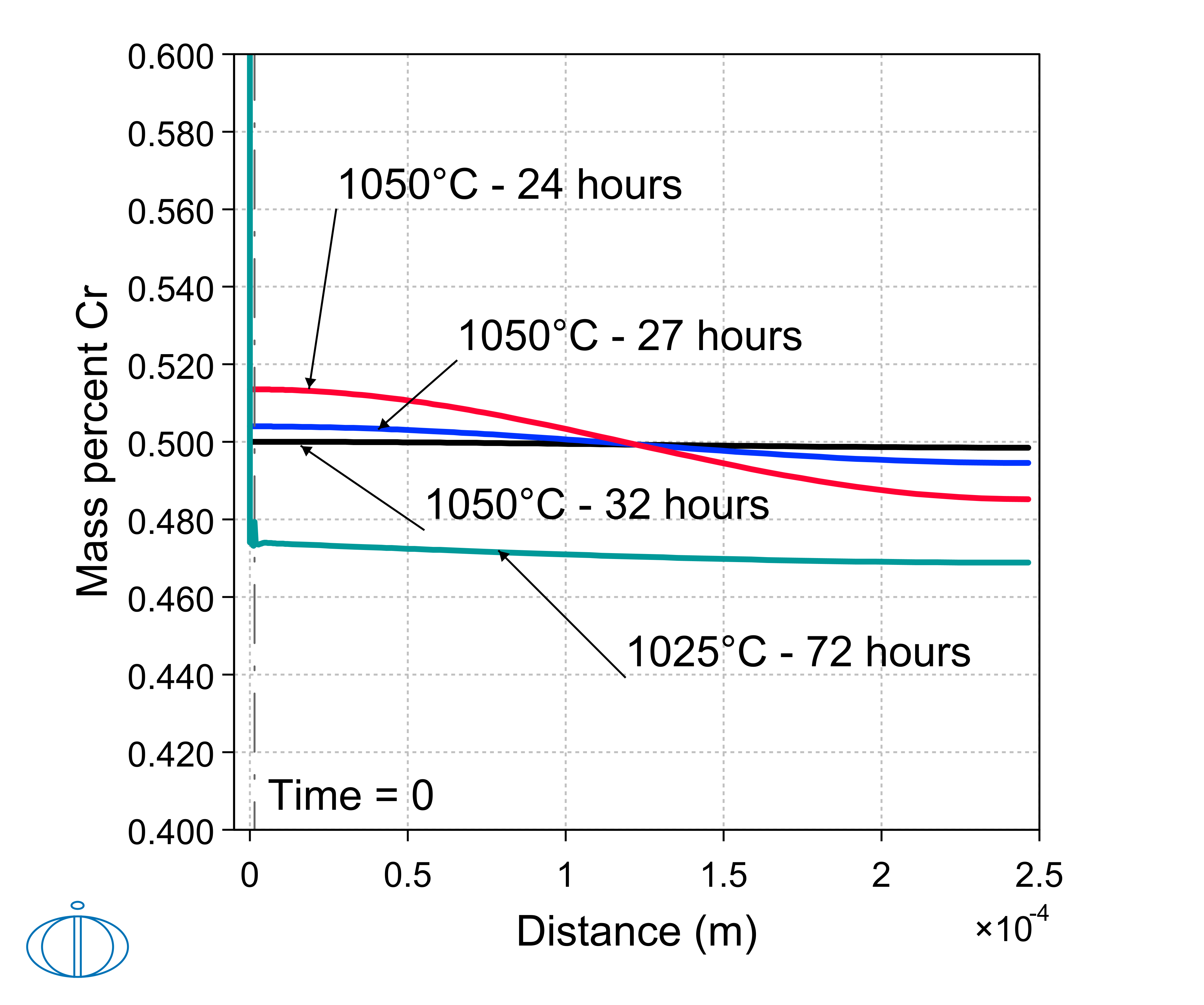}}
    \subfigure[\label{fig:1b}]{\includegraphics[width=0.31\linewidth]{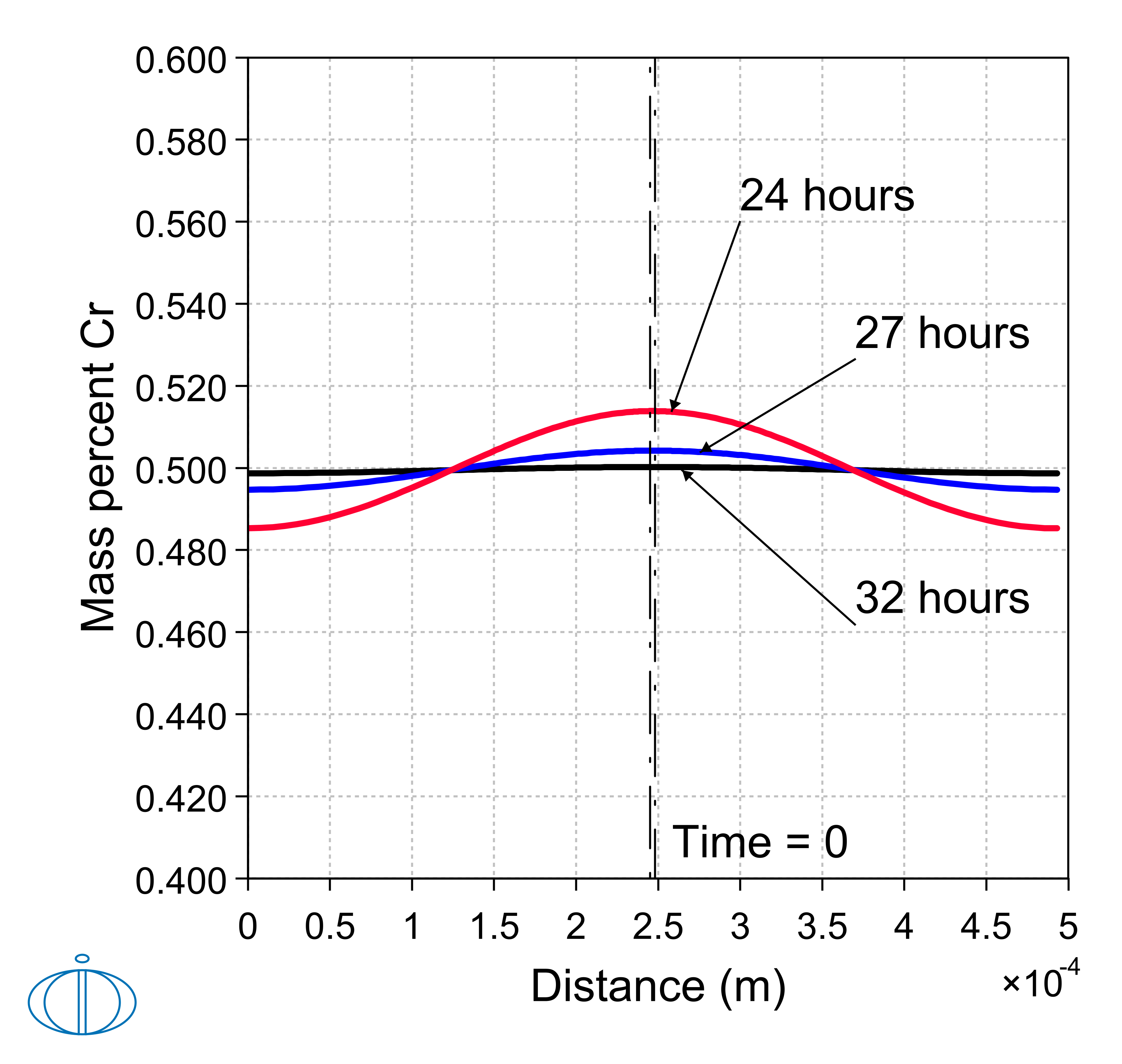}}
    \subfigure[\label{fig:1c}]{\includegraphics[width=0.31\linewidth]{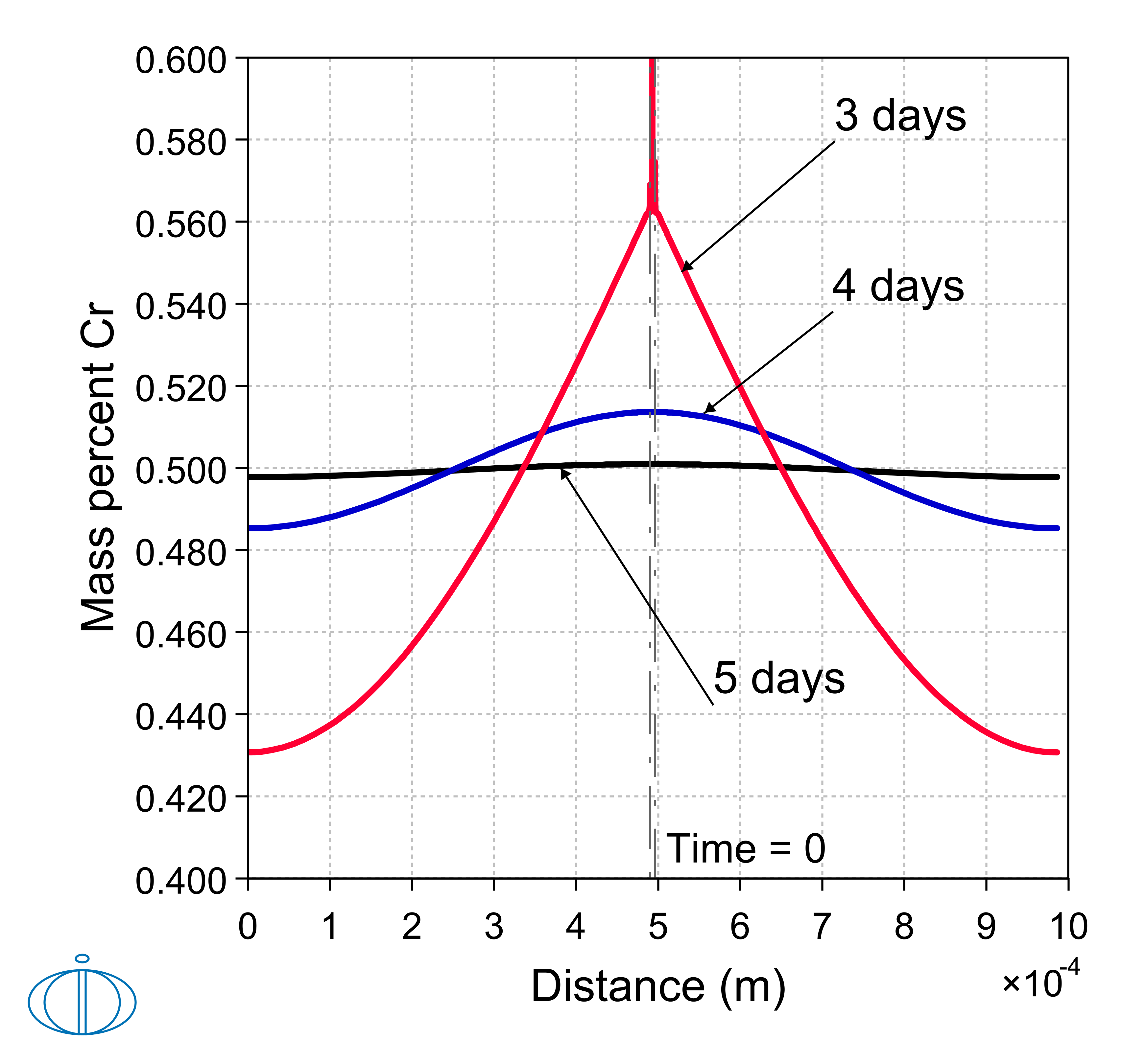}}
    \caption{Cr concentration profiles. Simulated Cr concentration profiles after solution heat treatment at $1050\,^{\circ}$C. 
    \subref{fig:1a} $1.5\;\si{\micro\meter}$ Cr in contact with $245\;\si{\micro\meter}$ Cu,  \subref{fig:1b} $3\;\si{\micro\meter}$ Cr in contact with $245\;\si{\micro\meter}$~Cu at each side, and \subref{fig:1c} $6\;\si{\micro\meter}$ Cr in contact with $490\;\si{\micro\meter}$~Cu at each side. The duration of solution heat treatment is indicated for each concentration profile. In \subref{fig:1a} the Cr concentration profile following solution heat treatment for 72~h at $1025\,^{\circ}$C is also shown for comparison. The logo of the DICTRA module is shown on the lower left corner of each panel.
    \label{fig:1}}
    \vspace{-0.5cm}
\end{figure}

In the first case, as shown in Fig.~\ref{fig:1a}, the single-sided layer configuration of $1.5\;\si{\micro\meter}$~Cr in contact with $245\;\si{\micro\meter}$~Cu is explored and homogenization can be achieved after 32~h. In Figs.~\ref{fig:1b} and \subref{fig:1c} the double-sided configuration is presented, namely a thicker Cr layer which is in contact with a Cu layer on each side. When Cr thickness increases from $1.5$ to $3\;\si{\micro\meter}$ and the thickness of the two Cu layers to $245\;\si{\micro\meter}$, homogenization is still achieved after the same time duration of 32~h. For the case of $6\;\si{\micro\meter}$ Cr in contact with two Cu layers of $490\;\si{\micro\meter}$ thick each, it will take 5 days for it to fully dissolve and obtain a homogeneous alloy composition of 0.5 wt\% Cr content.

From these results, it is seen that a substantial increase of the Cr thickness leads to much longer processing times that may become impractical. However, although a uniform Cr thickness at the few $\si{\micro\meter}$-level is challenging to obtain experimentally consistently, it is expected that small variations in the thickness of a thin Cr layer will not affect the alloy homogenization.

\subsubsection{Towards an electroformed CuCrTi alloy}
\label{sec:alloy2}
\begin{figure}
    \centering
    \subfigure[\label{fig:2a}]{\includegraphics[width=0.32\linewidth]{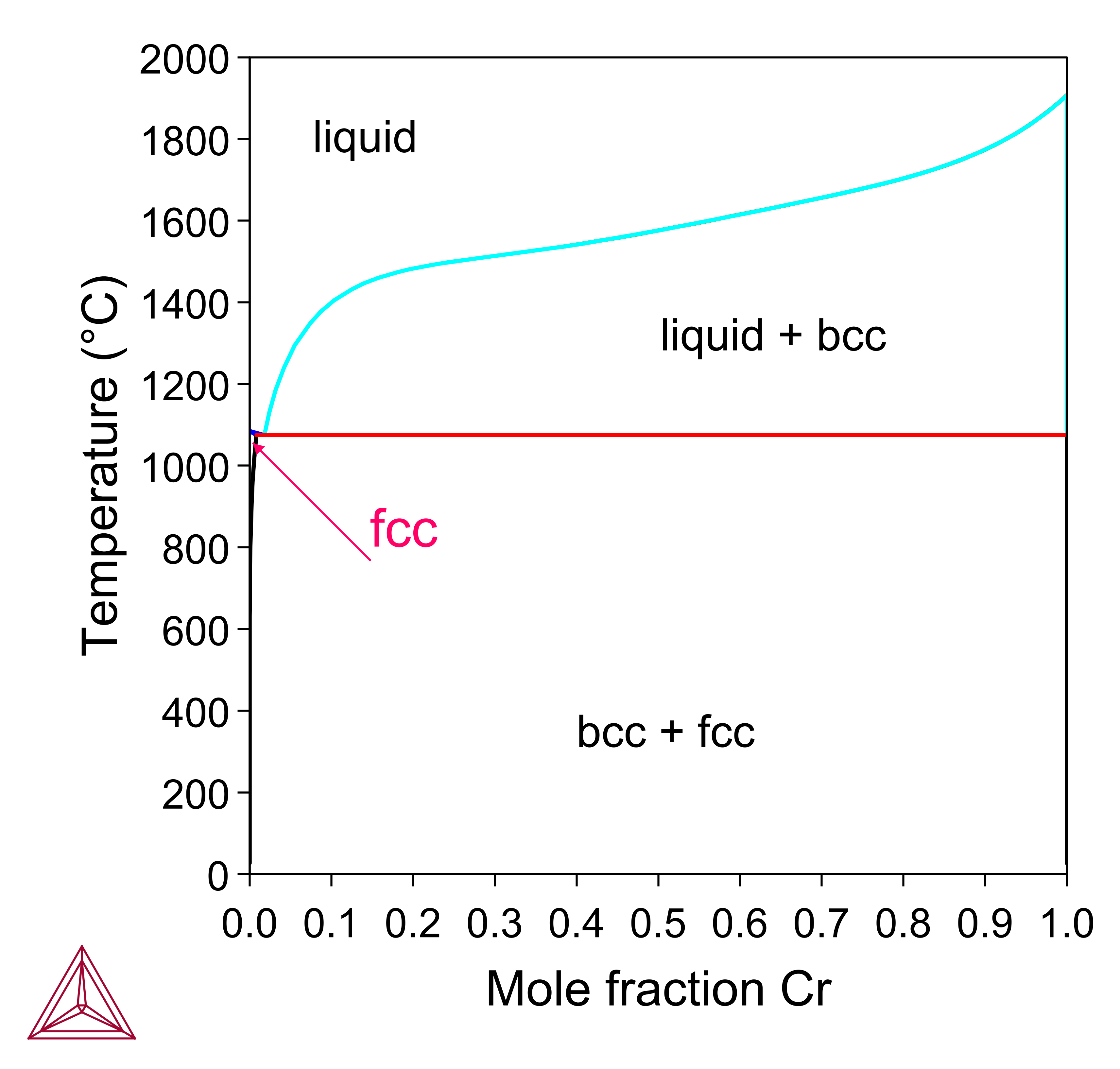}}
    \subfigure[\label{fig:2b}]{\includegraphics[width=0.32\linewidth]{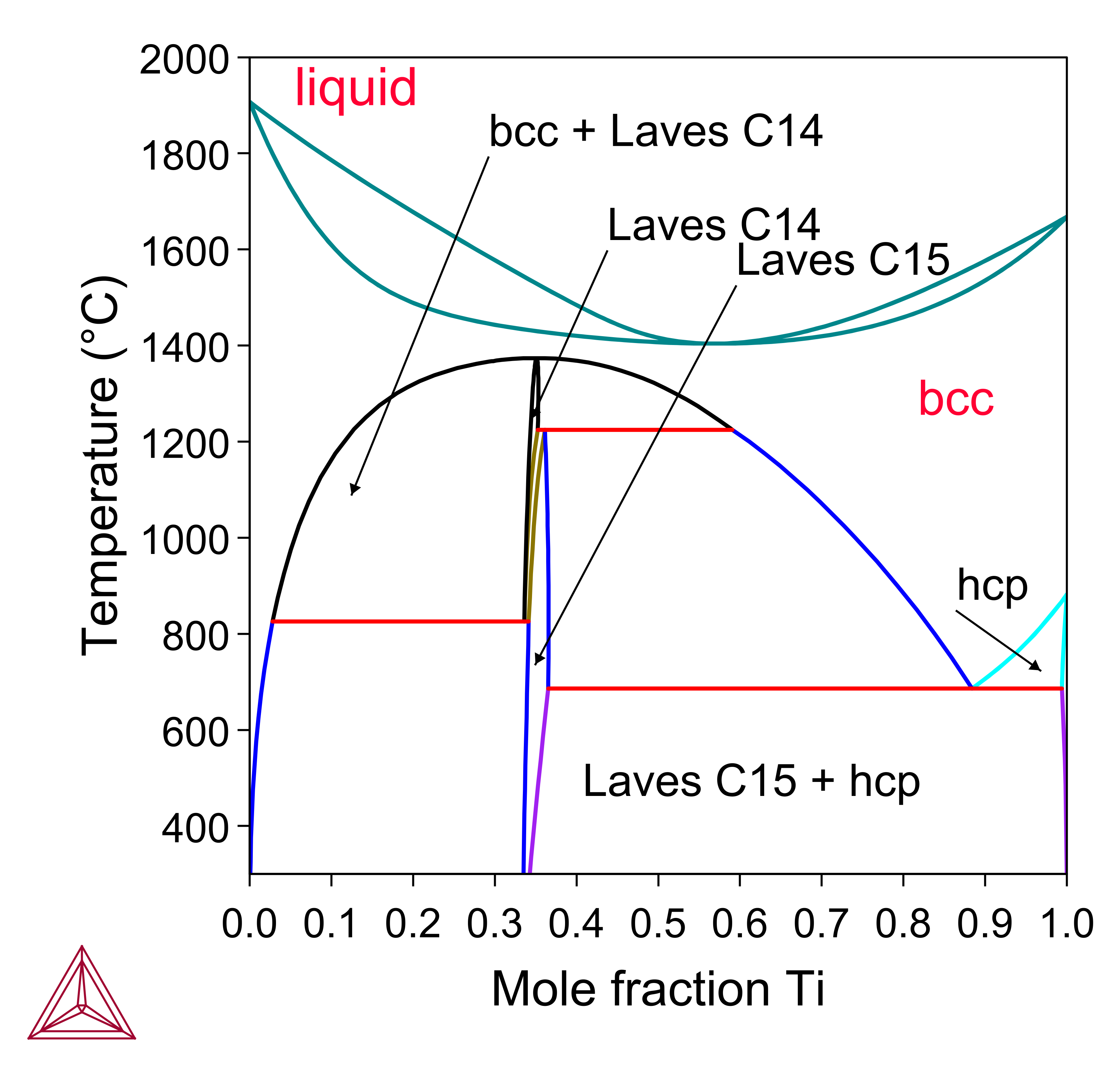}}
    \subfigure[\label{fig:2c}]{\includegraphics[width=0.32\linewidth]{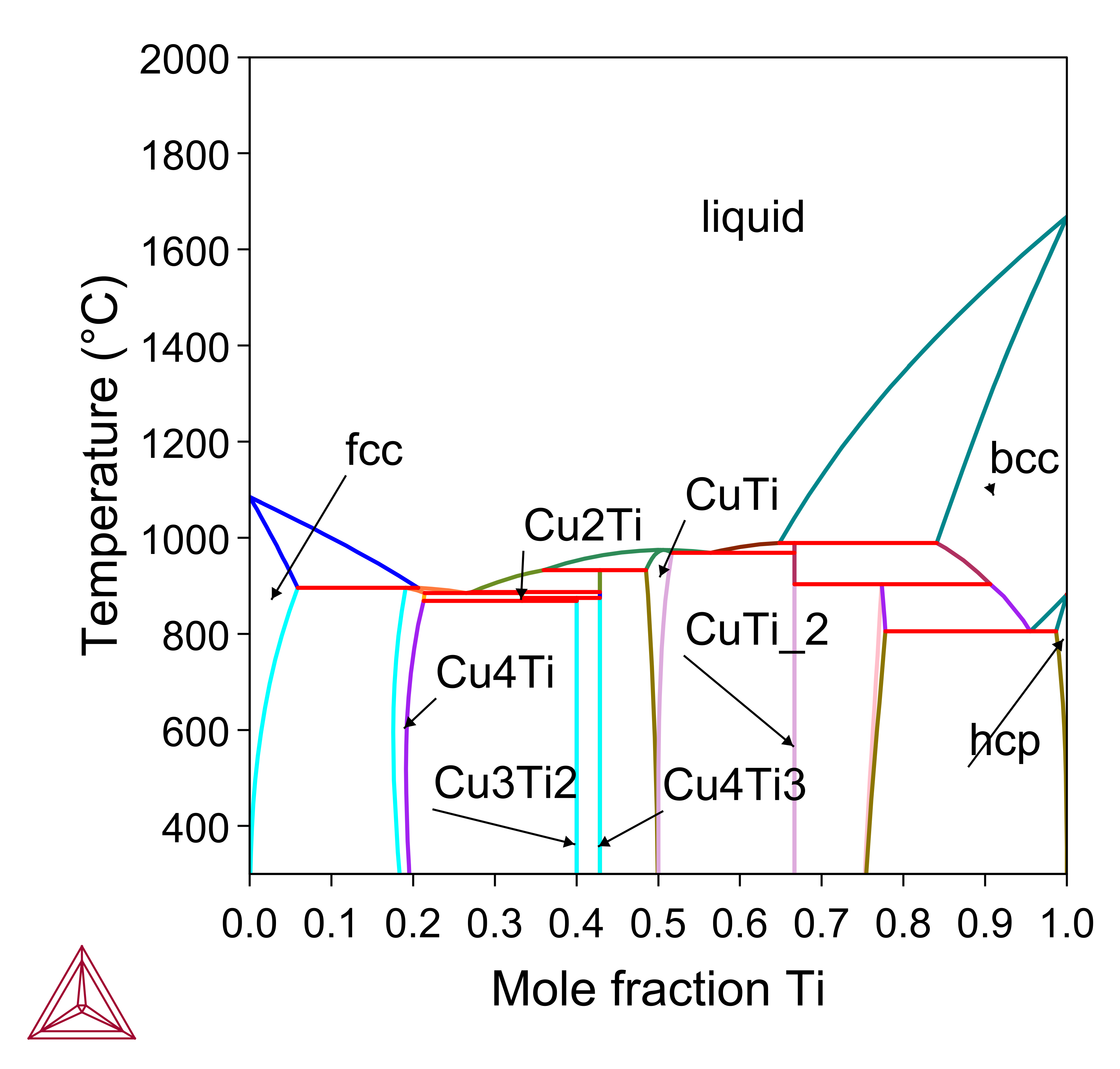}}
    \caption{Phase diagrams. Calculated phase diagrams for the \subref{fig:2a} Cu-Cr \subref{fig:2b} Cr-Ti and \subref{fig:2c} Cu-Ti systems. The TCHEA6 and MOBHEA3 High Entropy alloys Database is used~\cite{TCdatabase1, TCdatabase6}. Several phases are formed, including faced centered cubic (fcc), body centered cubic (bcc), and hexagonal close-packed (hcp). The logo of the Thermo-Calc software is shown on the lower left corner of each panel.\label{fig:2}}
    \vspace{-0.3cm}
\end{figure}

The phase diagrams of the Cu-Cr, Cr-Ti, and Cu-Ti systems are obtained from equilibrium calculations and are presented in Figs.~\ref{fig:2a}, \subref{fig:2b}, and~\subref{fig:2c}, respectively. It is noted that the Cr-Ti and Cu-Ti phase diagrams are much more complicated than that of the Cu-Cr, and consist of several phases when moving in the composition and temperature space between the two corresponding pure elements. In general, the thermal processing time for homogenization is substantially increased when multiple phase transformations occur. In addition, thermodynamic equilibrium is reached faster at higher temperatures. However, higher temperatures may need to be avoided, in particular in temperature-composition regions where the liquid phase is stable. 
Moreover, Fig.~\ref{fig:2c} indicates numerous intermetallics growing across the Cr-Ti composition range, making the homogenization of a Cu-Ti alloy challenging to achieve in a realistic duration of time in the case where EFCu is deposited on the Ti surface. For this reason, to achieve a CuCrTi alloy through electroformation, the initial diffusion of a thin Ti layer into a layer of Cr, rather than into a layer of Cu, is explored. 
The resulting CrTi alloy will subsequently be brought in contact with Cu layers. Moreover, phase transformation considerations, namely, avoiding melting, limit both the processing temperature and the target Ti concentration in the CrTi alloy. Apart from the temperature of $1050\,^{\circ}$C, which is the one used  for homogenization of Cu-0.5Cr alloy, also $1300\,^{\circ}$C and $1395\,^{\circ}$C are examined.

In the Cr-Ti system, Fig.~\ref{fig:2b}, a bcc single-phase microstructure is thermodynamically stable in the Cr-rich region. This region becomes narrower with decreasing temperature, in particular below $1200\,^{\circ}$C. For example, for processing at $1050\,^{\circ}$C only up to approximately 6.5 (wt\%) Ti is possible within Cr. In the adjacent phase-field a two-phase region is present, comprising bcc and Laves C14 phases. This means that at $1050\,^{\circ}$C, for example, when a thinner pure Ti layer is in contact with a thicker pure Cr layer and interdiffusion begins at the Cr/Ti interface, pure Ti, initially in bcc crystal structure, will first form the Laves C15 phase and then the Laves C14 phase until finally dissolving into the single-phase field of the Cr-rich bcc phase. Between approximately $1225\,^{\circ}$C and $1380\,^{\circ}$C, the pure Ti will first form the Laves C14 phase and then fully dissolve in the single-phase field of the Cr-rich bcc phase. For temperatures above approximately $1380\,^{\circ}$C, but below the melting point, pure Ti is in the same phase as pure Cr and no intermediate phases are formed. 

\begin{figure}
    \centering
    \subfigure[\label{fig:3a}]{\includegraphics[width=0.32\linewidth]{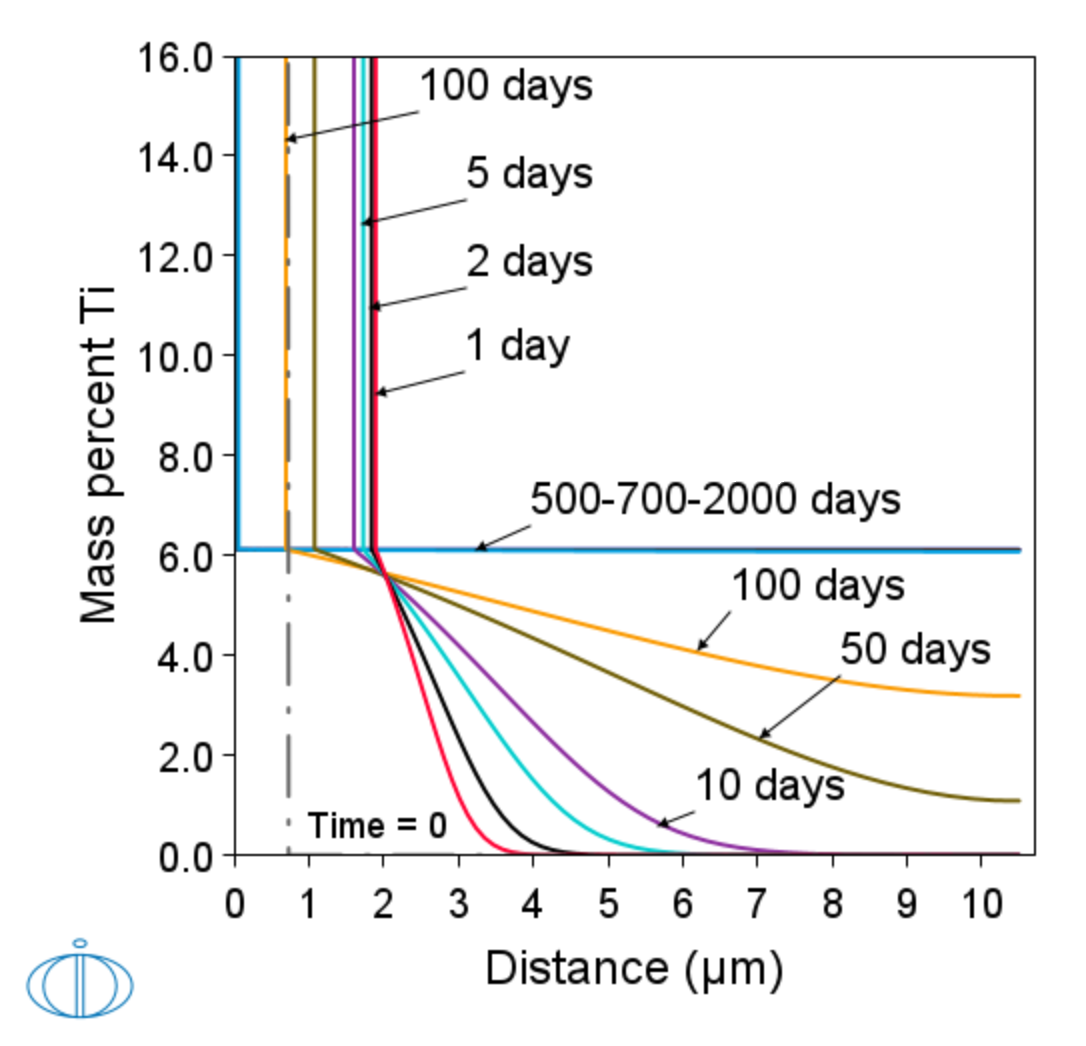}}
    \subfigure[\label{fig:3b}]{\includegraphics[width=0.32\linewidth]{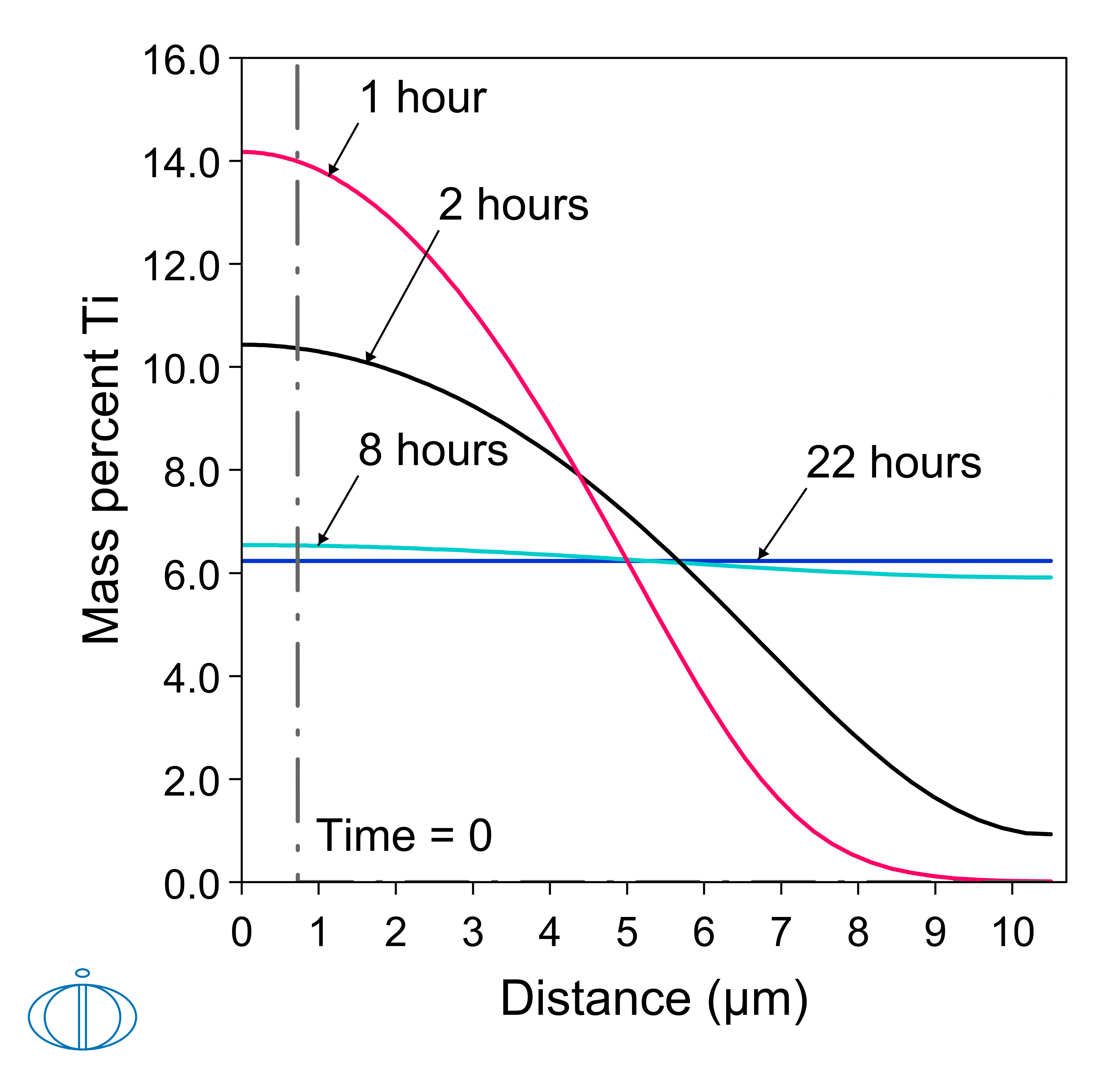}}
    \subfigure[\label{fig:3c}]{\includegraphics[width=0.32\linewidth]{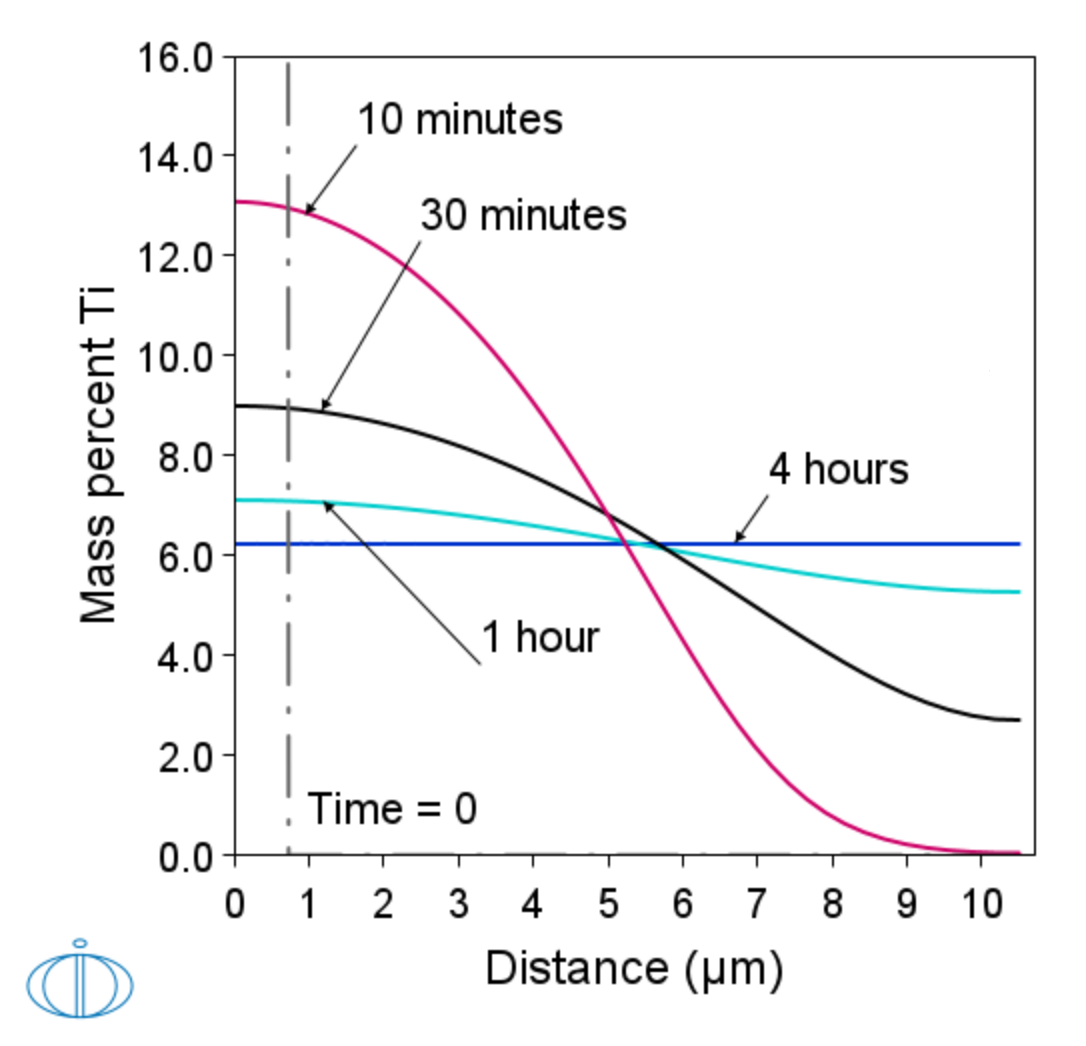}}
    \caption{Ti concentration profiles. Simulated Ti concentration profiles as a result of diffusion after solution heat treatment of 0.7~$\mu m$ Ti in contact with 9.8~$\mu m$ Cr. \subref{fig:3a} $1050\,^{\circ}$C,
    \subref{fig:3b} $1300\,^{\circ}$C and
    \subref{fig:3b} $1395\,^{\circ}$C. The duration of solution heat treatment is indicated for each concentration profile. The logo of the DICTRA module is shown on the lower left corner of each panel.
    \label{fig:3}}
    \vspace{-0.3cm}
\end{figure}

The challenge of achieving a homogenized CrTi alloy with a single-phase microstructure at the same processing temperature as that used for CuCr is highlighted in Fig.~\ref{fig:3a}. The simulation results of the solution heat treatment at $1050\,^{\circ}$C for up to 2000 days are presented in the case of $0.7\;\si{\micro\meter}$
Ti in contact with $9.8\;\si{\micro\meter}$ Cr. As discussed, in this case 
Laves phases C14 and C15 form according to the Cr-Ti phase diagram, shown in Fig.~\ref{fig:2b}, and their growth is driven mainly kinetically. As a result, both growth and dissolution of these phases proceed extremely slowly. This was also 
experimentally verified by the Cr-Ti diffusion couples reported in Ref.~\cite{flores2024experimental}.
In contrast, homogenization through solution heat treatment proceeds much faster at $1300\,^{\circ}$C and $1395\,^{\circ}$C, where  22~h and 4~h are required, respectively, for a final composition of Cr-6.23Ti (wt\%) in a $10.5\;\si{\micro\meter}$-thick layer.

\begin{figure}
    \centering
    \subfigure[\label{fig:4a}]{\includegraphics[width=0.43\linewidth]{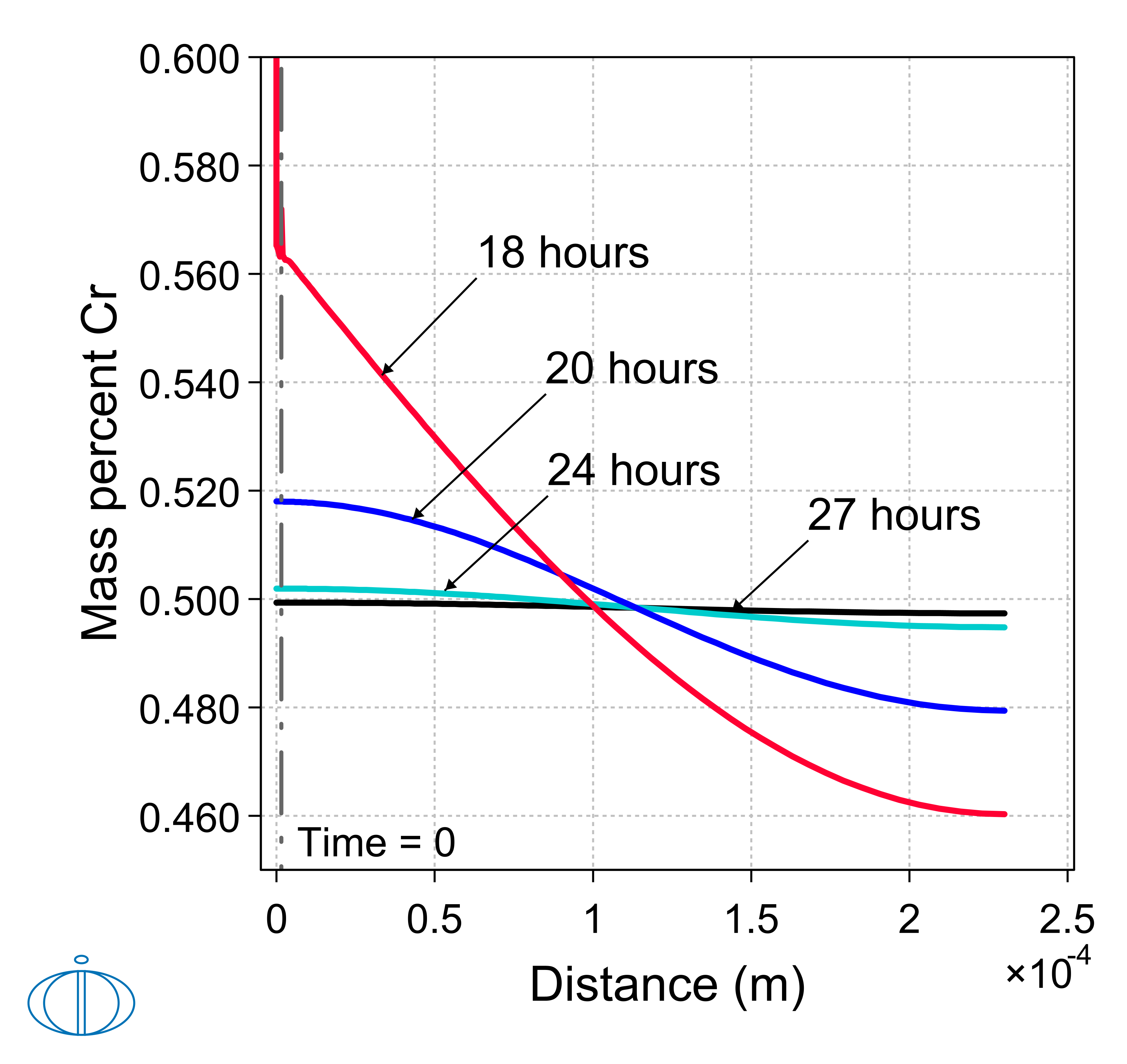}}
    \subfigure[\label{fig:4b}]{\includegraphics[width=0.43\linewidth]{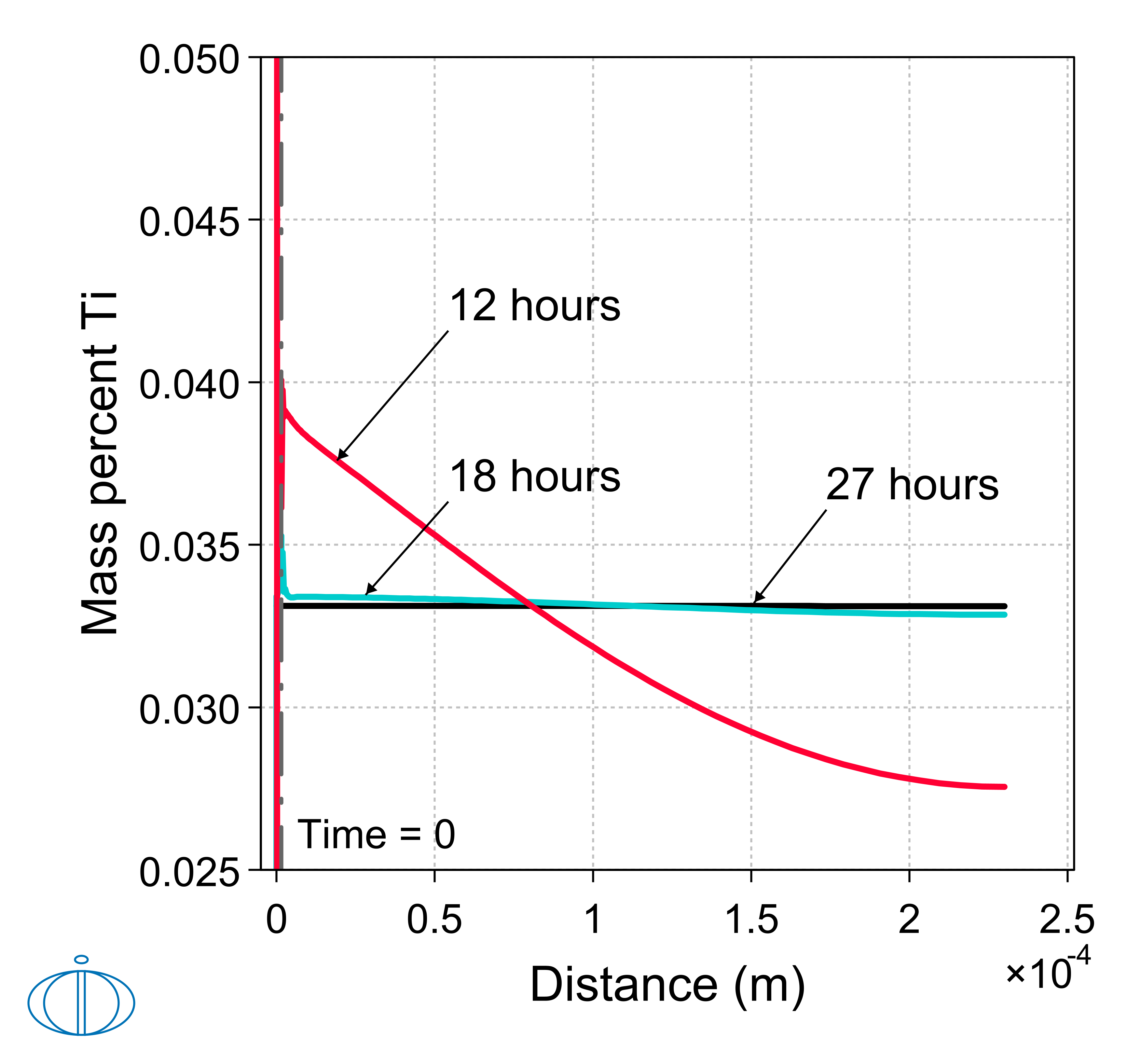}}
    \caption{Cr and Ti concentration profiles. Simulated concentration profiles as a result of diffusion after solution heat treatment at $1050\,^{\circ}$C of  1.5 $\mu m$ Cr-6.23Ti (in wt\%) in contact with 228.5~$\mu m$ Cu of  
    \subref{fig:4a} Cr and 
    \subref{fig:4b} Ti, resulting in homogenized alloy Cu-0.5Cr-0.032Ti composition, 230~$\mu m$ thick. The duration of solution heat treatment is indicated for each concentration profile. The logo of the DICTRA module is shown on the lower left corner of each panel.
    \label{fig:4}}
    \vspace{-0.3cm}
\end{figure}

The solution heat treatment at $1050\,^{\circ}$C of a $1.5\;\si{\micro\meter}$-thick homogenized Cr-6.23Ti (in wt\%) layer in contact with 
$228.5\;\si{\micro\meter}$ Cu layer leads to a homogenized Cu-0.5Cr-0.032Ti (in wt\%) alloy after 27~h. The resulting layer has a thickness of 
$230\;\si{\micro\meter}$.  The Cr and Ti concenration profiles during homogenization are shown in Figs.~\ref{fig:4a} and \subref{fig:4b}, respectively. 
Although fully homogenized concentration of both species is achieved at the same time, given that Ti content is substantially lower than that of Cr, Ti concentration is practically fully homogenized after 18~h. 
Full homogenization of the Cu-0.5Cr-0.032Ti is predicted to be achieved  after 27 hours of solution heat treatment at $1050\,^{\circ}$C, faster, compared to Cu-0.5Cr, which requires 32~h at the same temperature, as shown in Fig.\ref{fig:1}.

\subsection{Case studies: DarkSPHERE and XLZD}
The impact of this work is presented for two complementary next-generation dark matter direct detection experiments, namely DarkSPHERE and XLZD, which are utilizing completely different detection techniques. It is, thus, expected that the majority of future rare-event search experiments would benefit from the presented developments.

\subsubsection{The DarkSPHERE detector}
DarkSPHERE~\cite{NEWS-G:2023qwh} is a proposed next-generation experiment designed to search for light dark matter using spherical proportional counters~\cite{Giomataris:2008ap, Nikolopoulos_Knights_2025}. The detector comprises a grounded spherical metallic vessel filled with a suitable gas mixture and a central read-out structure equipped with anodes of approximately~$1\;\si{\milli\meter}$ in radius. The read-out structure is mounted on a grounded metallic rod, which also serves to shield the wire that delivers positive voltage to the anode and transmits the signal for read-out. The electric field inside the vessel decreases approximately as $1/r^2$ with distance from the centre, thereby dividing the gas volume into drift and avalanche regions. When particles interact within the gas volume, they can ionize atoms, producing electrons that drift towards the central anode. As these electrons approach within about $100\,\si{\micro\meter}$ of the anode, the electric field becomes strong enough to initiate an avalanche, thereby amplifying the signal. Earlier detectors such as {\sc SEDINE}~\cite{NEWS-G:2017pxg} and {\sc S140}~\cite{NEWS-G:2022kon} have demonstrated the viability of the technique and have established leading constraints in the search for sub-GeV particle dark matter candidates~\cite{NEWS-G:2024jms,NEWS-G:2017pxg}. 

The proposed DarkSPHERE detector incorporates a number of recent developments in spherical proportional counter read-out~\cite{Giganon:2017isb, Katsioulas:2018pyh, Giomataris:2020rna, Katsioulas:2022cqe, Herd:2023hmu} and in EFCu fabrication~\cite{NEWS-G:2020fhm}, to extend this approach by constructing a large-scale fully-electroformed underground spherical proportional counter with a diameter of $300\;\si{\centi\meter}$ at the Boulby Underground Laboratory. The spherical proportional counter will be complemented by a modular pure-water shield.
This substantial increase in active volume, combined with ultra-low-background Cu fabrication and the  water-based shielding, is expected to significantly boost sensitivity~\cite{NEWS-G:2023qwh}.

In the baseline design, DarkSPHERE is to initially operate with a helium-isobutane (90\%:10\%) gas mixture at a pressure of $5\;\si{\bar}$, providing kinematic sensitivity to light dark matter. With a total target mass of approximately of $27\,\si{\kilogram}$, it is envisaged to reach the so-called ``neutrino fog''~\cite{OHare:2021utq} for dark matter-nucleon scattering cross-sections at low masses. The choice of target gas mixture further enables sensitivity to a broad range of dark matter-nucleus and dark matter-electron interactions, thanks to the natural carbon-13 abundance and single ionization electron threshold capabilities~\cite{NEWS-G:2019lqz}. For dark matter electron scattering, in particular, it has been shown that molecules such as i-C$_4$H$_{10}$ provide enhanced sensitivity thanks to the weak bond of outer shell electrons~\cite{Hamaide:2021hlp}.

EFCu meets the radiopurity, electrical conductivity, and mechanical requirements of DarkSPHERE. However, the ductility of Cu dictates the cathode shell thickness and thus the highest possible detector operating pressure for structural stability. EFCu fabrication proceeds slowly, with approximately $1\;\si{\milli\meter}$ growing per month~\cite{NEWS-G:2020fhm}. Thus, for a practical detector construction time, the experiment is limited to a thickness of $1\;\si{\centi\meter}$. For a thin-walled spherical shell, such as the DarkSPHERE vessel, the stress $\sigma$ under internal pressure is given by $\sigma = \frac{p r}{2 t}$,  where $r$ and $t$ are the radius and wall thickness of the spherical shell  subjected to a uniform internal pressure $p$ above the external atmospheric pressure. This sets the mentioned $5\;\si{\bar}$ limit in operating pressure, when safety tolerance is also considered.

The use of a high-purity alloy would result in a stronger detector vessel without compromising radiopurity. A double of the strength with respect to EFCu would enable detector operation at a gas pressure higher by a factor of two. This halves the data-collection time for the experiment to achieve the exposure required for its physics goals. As a result, for the same research programme duration, a second dataset of similar exposure can be collected with a different gas mixture. 

\begin{figure}[h] 
\centering
\subfigure[\label{fig:limitsSI}]{\includegraphics[width=0.32\columnwidth]{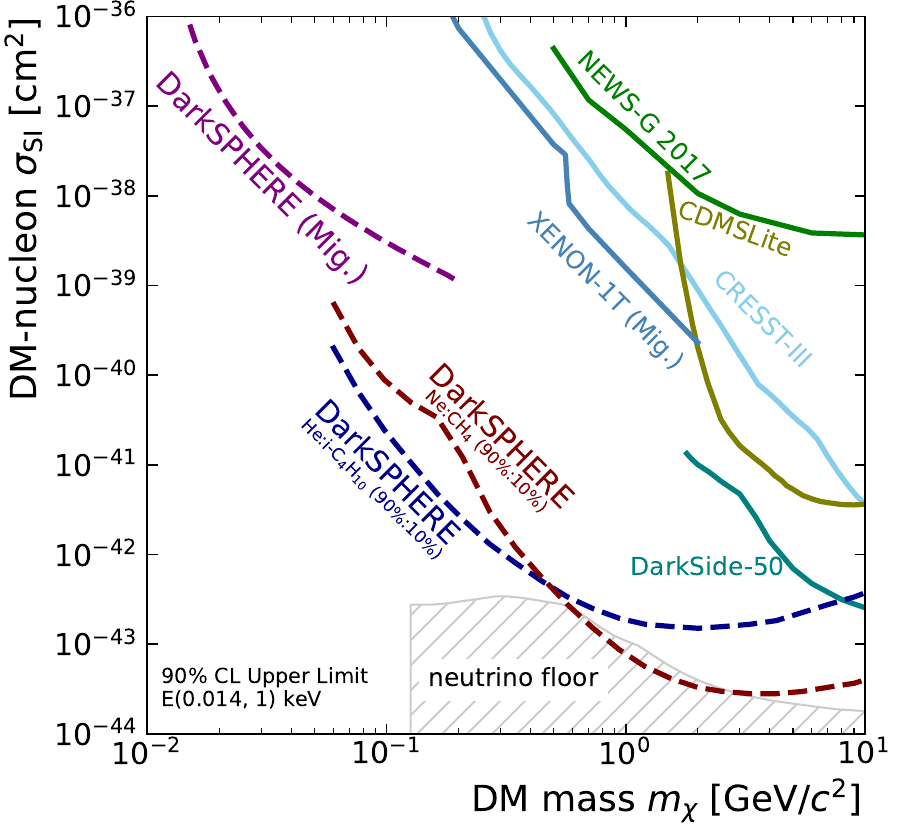}}
\subfigure[\label{fig:limitsSDp}]{\includegraphics[width=0.32\columnwidth]{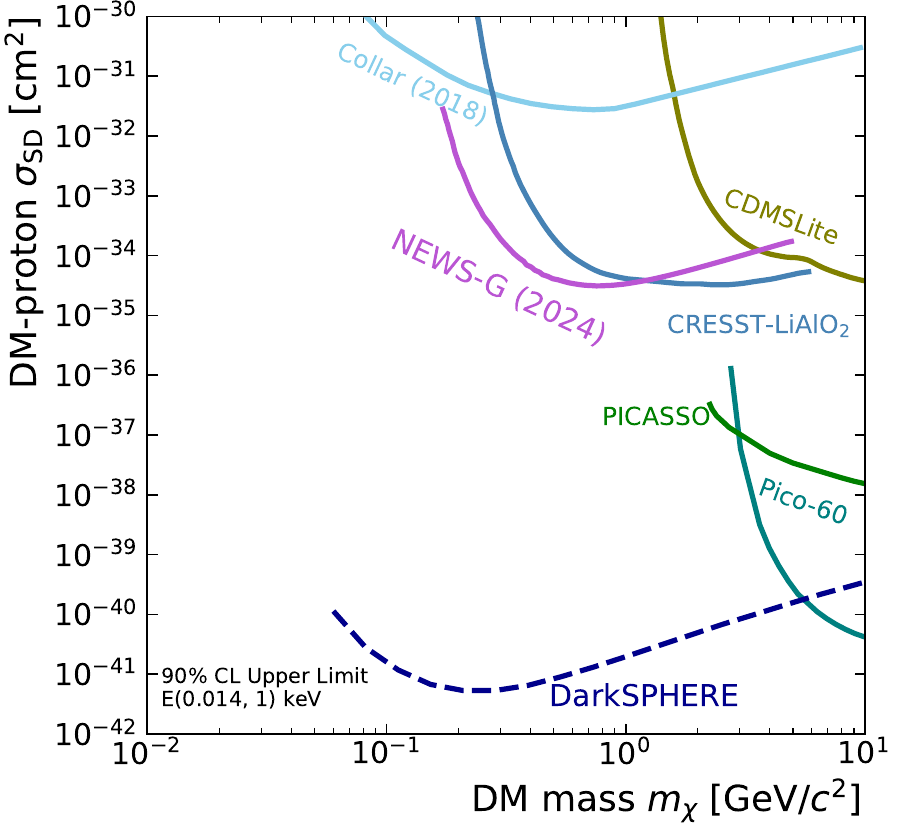}}
\subfigure[\label{fig:limitsSDn}]{\includegraphics[width=0.32\columnwidth]{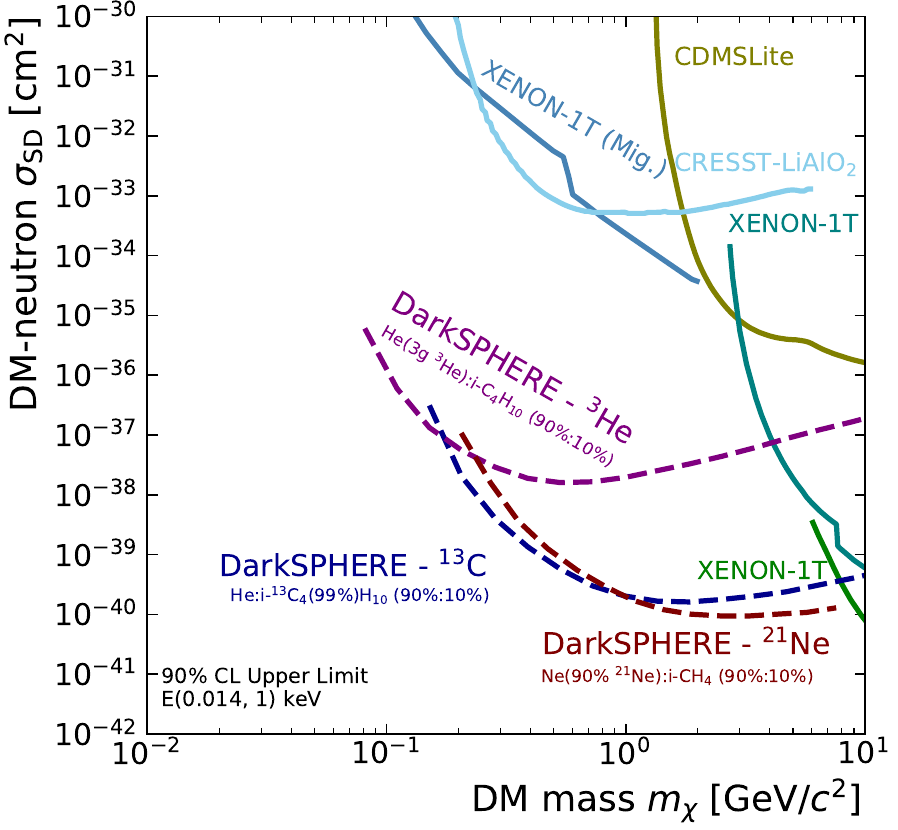}}
  \vspace{-0.3cm}
  \caption{DarkSPHERE sensitivity. Expected sensitivity of DarkSPHERE for \subref{fig:limitsSI}~spin-independent dark matter-nucleon interaction cross-section, including the enhancement due to the Migdal effect (Mig.) from Ref.~\cite{NEWS-G:2023qwh}, and spin-dependent \subref{fig:limitsSDp}~dark matter-proton and \subref{fig:limitsSDn}~dark matter-neutron interaction cross-section. All assume a nominal exposure of 300 days and a gas at a pressure of $5\;\si{\bar}$~\cite{NEWS-G:2023qwh}. The spin-dependent proton sensitivity includes the contribution from both the He:i-C$_4$H$_{10}$ and the Ne:CH$_4$ data taking campaigns. The sensitivity in \subref{fig:limitsSDn} comes from using Ne:CH$_4$ isotopically enriched in $^{21}$Ne and i-C$_4$H$_{10}$ that is isotopically enriched in $^{13}$C to 99\% while in Ref.~\cite{NEWS-G:2023qwh} the $^{13}$C natural abundance was considered. Also shown is the sensitivity where $5\;\si{\gram}$ of helium are replaced with $^3$He. Existing constraints are summarised both for spin-independent~\cite{Agnese:2018gze,Abdelhameed:2019hmk,Agnes:2018ves,NEWS-G:2017pxg,Aprile:2019jmx} and spin-dependent~\cite{SuperCDMS:2017nns,Collar:2018ydf,CRESST:2022dtl,Behnke:2016lsk,PICO:2019vsc,Aprile:2019jmx, CRESST:2022dtl} searches.
  \label{fig:limits}}
  \vspace{-0.3cm}
\end{figure}

The potential physics benefits of such a development are shown in Figs.~\ref{fig:limitsSI}, \subref{fig:limitsSDp}, and \subref{fig:limitsSDn}, where the expected DarkSPHERE sensitivity for spin-independent dark matter-nucleon, spin-dependent dark matter-proton, and spin-dependent dark matter-neutron interaction cross-sections are summarized. Exposure of approximately $8\;\si{\tonne\cdot days}$, assuming operation with the baseline He:i-C$_4$H$_{10}$ (90\%:10\%)  mixture at $5\;\si{\bar}$, would require 300 days~\cite{NEWS-G:2023qwh}. The use of a high-purity CuCr alloy with double the strength over EFCu, would allow operation at $10\;\si{\bar}$ and reduce the required time to 150 days. This would enable another physics run in the same initial planned research program duration. Building on previous NEWS\=/G experience of using Ne:CH$_{4}$~\cite{NEWS-G:2017pxg}, for example, would not require additional research and development for detector operation. A subsequent operation for 150 days with Ne:CH$_{4}$ (90\%:10\%) at $10\;\si{\bar}$ would allow a second dataset to be collected. The extended physics reach using this second dataset is shown in Fig.~\ref{fig:limitsSI} for the spin-independent dark matter-nucleon interaction and in \subref{fig:limitsSDn} for the the spin-dependent dark-matter-neutron interaction through the use of Ne, which is 90\% isotopically enriched in $^{21}$Ne. The use of Ne allows an additional parameter space to be explored at higher DM candidate masses,  while the use of $^{21}$Ne would provide a second target nucleus to independently probe this parameter space with similar sensitivity. In Fig.~\ref{fig:limitsSDn} the use of i-C$_4$H$_{10}$ that is isotopically enriched in $^{13}$C to 99\% is assumed, while in Ref.~\cite{NEWS-G:2023qwh} the natural abundance of $^{13}$C was considered. Moreover, also shown in Fig.~\ref{fig:limitsSDn} is the sensitivity where $5\;\si{\gram}$ of helium is replaced by $^3$He.

\subsubsection{The XLZD detector}

XLZD builds on the legacy of leading liquid xenon experiments, and in particular of LUX-ZEPLIN (LZ)~\cite{PhysRevLett.131.041002} and XENON-nT~\cite{XENON:2024wpa,XENON:2025vwd}, as well as extensive research and development conducted by the DARWIN Collaboration~\cite{DARWIN:2023uje,DARWIN:2020jme}. The XLZD Collaboration is pursuing a next-generation dual-phase liquid xenon time projection chamber (TPC)~\cite{Aalbers:2022dzr, XLZD:2024nsu} that will be at the heart of a broad multi-physics program. 
The nominal design~\cite{XLZD:2024nsu} features a cylindrical TPC 
of $2.98\;\si{\meter}$ diameter and $2.97\;\si{\meter}$ height, where an active mass of $60\;\si{\tonne}$ of liquid xenon is housed in a double-walled cryostat. 

Inside the TPC, the large active mass of liquid xenon -- with a small gas region at the top -- will be viewed by arrays of low-background photomultiplier tubes, enabling the measurement of primary interaction scintillation and light produced by electrons drifted to the gaseous region to reconstruct event energies and positions with high accuracy. The detector will be surrounded by extensive shielding, utilizing both passive layers and an instrumented active veto to further suppress external backgrounds.  This apparatus is designed to probe new regions of parameter space in direct dark matter searches, targeting spin-independent dark matter-nucleon cross-sections below $\num{E-48}\;\si{\centi\meter\squared}$ for a WIMP of $40\;\si{\giga\eV}$  mass. Additionally, XLZD aims for a 3$\sigma$ observation potential of neutrinoless double beta decay of $^{136}$Xe at a half-life of up to $\num{5.7E27}\;\si{years}$. To achieve its scientific objectives, XLZD will require substantial advances in the radiopurity of detector materials compared to current experiments.

Materials in contact with the xenon target -- especially those used for the cryostat vessel -- are critically important.
Beyond radiopurity, the cryostat also serves as a pressure vessel, imposing additional demands on mechanical strength. Previous large-scale experiments have used Ti and stainless steel for their mechanical strength; however, achieving further reductions in background, particularly that resulting from radon emanation, is essential for XLZD to reach its physics goals~\cite{LZ:2017iwn}.
Although efforts are ongoing to explore chemical etching, surface coatings~\cite{Jorg:2022spz}, and high-purity copper electroformation~\cite{Hoppe:2014nva,NEWS-G:2020fhm}, an attractive alternative would be a high-purity, mechanically robust alloy. A radiopure alloy with superior strength to Cu, but without compromising on cleanliness, would be a promising substitute for conventional materials in the cryostat and could also be employed in supporting structures within the experiment. Commercially available pure Ti and 316 stainless steel, typically, have yield strengths of $170-310$ and $240\;\si{\mega\pascal}$, respectively~\cite{structuralalloys,10.31399/asm.hb.v01.9781627081610}. 

In the comprehensive mechanical analysis of additive-free EFCu for MAJORANA Demonstrator a
typical yield strength of $80\;\si{\mega\pascal}$ and a hardness of  $(80\pm 6)$~HV was reported~\cite{osti_1039850}. This 
represents a factor of 2.1-3.8 and 3 lower yield strength compared to Ti and steel, respectively, due to the EFCu fine-grained microstructure.
For industrial Cu-0.5Cr (wt\%) alloy from casting, a measured yield strength of 
$(429\pm22)\;\si{\mega\pascal}$ and a hardness of  $160$~HV post thermal processing has been reported~\cite{HUANG2021102378}. It is noted that computational thermodynamics was implemented in that case to optimise the thermal processing parameters. It is anticipated that further mechanical enhancement can be achieved with refined strengthening techniques. Thus, the electroformed CuCr and CuCrTi alloys discussed hold promise to become an attractive low-background alternative to Ti and steel.

\section{Methods}
To date, the development of radiopure alloys has relied on intensive experimentation and iterative approaches to achieve improvements in mechanical properties. This trial-and-error process can be significantly accelerated with a better understanding of the underlying microstructure evolution and the resulting physical properties. This would enable the design of application-specific alloys tailored to meet both mechanical and radiopurity requirements.

Because the synthesis and mechanical enhancement of radiopure CuCr alloys are primarily achieved via thermal processing, predictive tools based on thermodynamic and kinetic modeling can be employed. The CALculation of PHAse Diagrams (CALPHAD)~\cite{lukas2007computational, SPENCER20081, kattner2016calphad} approach -- also known as computational thermodynamics -- provides a robust and widely used framework for simulating phase diagrams, phase stability, and microstructural evolution in alloys~\cite{national2008integrated,lu2014computational,LUO20156,de2019new, li2021calphad}. The approach of Integrated Computational Materials Engineering (ICME)~\cite{LUO20156, HUANG2021102378, Spathara:2025bfw} 
enables accelerated alloy development with targeted properties, and is applicable for manufacturing low-radioactivity materials~\cite{Spathara:2025bfw}.

A methodology -- based on CALPHAD-guided design strategies -- has recently been suggested to achieve homogenized CuCr alloy compositions with single-phase microstructures~\cite{Spathara:2025bfw}. These simulations were validated by direct comparison with available experimental investigations~\cite{Vitale:2021xrm}, and subsequently optimized thermal processing parameters designed to maximize the mechanical enhancement of radiopure CuCr alloys were proposed~\cite{Spathara:2025bfw}.

To predict composition profiles during the homogenization process, DICTRA 1D simulations were employed~\cite{andersson2002thermo}. The CALPHAD-type databases ``Thermo-calc Software TCHEA6 and MOBHEA3 High Entropy alloys Databases''~\cite{TCdatabase1, TCdatabase6} for thermodynamic and kinetic description of the Cu-Cr system, respectively, are used, similar to Ref.~\cite{Spathara:2025bfw}.
In these simulations, the diffusion equations are solved at the interface of the two regions: Cr and Cu layers resulting from sequential electroplating, which initially possess different crystal structures.  As homogenization proceeds in a diffusion-controlled manner, i.e. interdiffusion of Cu and Cr occurs from each of the two regions, and phase transformation occurs at the Cr/Cu interface. Simulations take place in a volume-fixed frame of reference~\cite{andersson1992models,spathara2018study}.

It is noted that, in the case of full homogenization, the simplifications of 1D modeling, e.g. neglecting roughness. are not expected to be important. The DICTRA 1D model has been validated on the same scale for a Ni-based superalloy, where Ni vapor was deposited on the superalloy surface for 24 hours~\cite{spathara2018study}. In the case where full homogenization is not achieved, differences between the 1D model and the actual materials may arise at the interface. This is discussed in Ref.~\cite{Spathara:2025bfw}, where calculations are compared with the experimental data. 

The solution heat treatment promotes homogenization of the alloy composition in the single-phase region, preparing the material for rapid cooling (quenching). At the temperature where the solution heat treatment takes place, Cu and Cr have different crystal structures. According to the calculated Cu-Cr phase diagram using the thermodynamic database TCHEA6, which is shown in Fig.~\ref{fig:2a}, Cr is in bcc phase, while Cu is in fcc phase. At sufficiently high temperatures, a single fcc phase becomes thermodynamically stable, in the Cu-rich area, close to and below the Cu melting point, allowing Cr atoms to diffuse into the Cu lattice until a uniform solid solution is achieved. This supersaturated solid solution is retained through quenching. The atomic size mismatch between Cu and Cr creates lattice distortions, which hinder dislocation motion and thereby strengthen the matrix. The type and concentration of solute atoms are both critical in determining these strengthening effects~\cite{zhang2019review, yang2023recent}.

CuCrTi alloys are found to have an increased hardness at levels comparable to those of a mild steel, i.e. 160 to 220~HV, fabricated from casting and thermally processed~\cite{HUANG2021102378}. We explore thermal processing towards homogenized, radiopure CuCrTi alloys from electroformation. The DICTRA simulation setup is similar to that described for the simulation of thermal processing of CuCr alloys. In this case, additional intermediate phases are possible beyond the bcc phases of the pure Cr and pure Ti, and the fcc phase of the pure Cu.

\section{Data availability}
Data supporting the findings of this study are available from the corresponding author on a reasonable request.

\section{References}
\bibliographystyle{JHEP}
\bibliography{bibliography}

\begin{acknowledgments}
Support from the UKRI-STFC (No. ST/W000652/1, No. ST/X005976/1) 
and the UKRI (No. EP/X022773/1) is acknowledged.
The support of the Deutsche Forschungsgemeinschaft (DFG, German Research Foundation) under Germany’s Excellence Strategy -- EXC 2121 “Quantum Universe”-390833306 is acknowledged.
\end{acknowledgments}

\section{Author contributions}
The content of this article was developed through direct
collaboration, discussions, and exchange of ideas among all authors
(DS, PK, KN). All authors contributed to the preparation of the
article through their respective expertise.

\section{Declaration of competing interest}
The authors declare no competing interests.

\clearpage

\end{document}